\documentclass[twocolumn,prb,prbprintnumbers,showpacs,fleqn,superscriptaddress]{revtex4}
\usepackage{epsfig}
\usepackage{graphicx}
\usepackage{amsfonts}
\usepackage{bm}
\usepackage{color}

\newcommand{\ct}{\cite}
\newcommand{\bi}{\bibitem}

\newcommand{\ket}{\rangle}
\newcommand{\non}{\nonumber}

\newcommand{\be}{\begin{equation}}
\newcommand{\ee}{\end{equation}}
\newcommand{\ba}{\begin{eqnarray}}
\newcommand{\ea}{\end{eqnarray}}

\begin{document}
\title{Oscillating fidelity susceptibility near a quantum multicritical point}

\author{Victor Mukherjee}
\email{victor@iitk.ac.in}
\affiliation{Department of Physics, Indian Institute of Technology Kanpur, Kanpur 208 016, India}
\author{Anatoli Polkovnikov}
\email{asp@buphy.bu.edu}
\affiliation{Department of Physics, Boston University, 590 Commonwealth Avenue, Boston, Massachusetts 02215, USA}
\author{Amit Dutta}
\email{dutta@iitk.ac.in}
\affiliation{Department of Physics, Indian Institute of Technology Kanpur, Kanpur 208 016, India}

\begin{abstract}
We study scaling behavior of the geometric tensor $\chi_{\alpha,\beta}(\lambda_1,\lambda_2)$ and the fidelity susceptibility $(\chi_{\rm F})$ in the
vicinity of a quantum multicritical point (MCP) using the example of a transverse $XY$ model. We show that the behavior of the geometric tensor
(and thus of $\chi_{\rm F}$) is drastically different from that seen near a critical point. In particular, we find that is highly non-monotonic
function of $\lambda$ along the generic direction $\lambda_1\sim\lambda_2 = \lambda$ when the system size $L$ is bounded between the shorter and longer correlation
lengths characterizing the MCP: $1/|\lambda|^{\nu_1}\ll L\ll 1/|\lambda|^{\nu_2}$, where $\nu_1<\nu_2$ are the two correlation length exponents
 characterizing the system. We find that the scaling of the maxima of the components of $\chi_{\alpha\beta}$ is associated with emergence of
  quasi-critical points at $\lambda\sim 1/L^{1/\nu_1}$, related to the proximity to the critical line of finite momentum anisotropic transition.
 This scaling is  different from that in the thermodynamic limit $L\gg 1/|\lambda|^{\nu_2}$, which is determined by the conventional critical exponents.
 We use our results to calculate the defect density following a rapid quench starting from the MCP and show that it exerts a step-like behavior for
 small quench amplitudes. Study of heat density and diagonal entropy density also show signatures of quasi-critical points.

\end{abstract}
\maketitle

\section{Introduction and main results}
Fidelity  and fidelity susceptibility\ct{zanardi06,venuti07,li09, gurev08, shigu08} ($\chi_{\rm F}$) are information theoretic measures of continuous quantum phase transitions (QPT) . It is well known that a QPT \ct{sachdev99,dutta96}, occurring at absolute zero temperature, indicates a change in symmetry of the ground state of a many body quantum system. Fidelity measures the overlap between two neighboring ground states in the parameter space of a quantum Hamiltonian and shows a dip at the quantum critical point (QCP) where the overlap between two neighboring ground states is minimum. At the same time, the fidelity susceptibility, which is a quantitative measure of the rate of change of the ground state of a system under an  infinitesimal variation of a parameter of the Hamiltonian, usually attains the maximum value and  often diverges with the system size in the thermodynamic  limit. The scaling behavior of $\chi_{\rm F}$   with the system size at the QCP and with respect to the deviation from the QCP  is given in terms of the associated quantum critical exponents.

Following the Kibble Zurek (KZ)  predictions \ct{zurek05, polkovnikov05, aeppli10, polkovnikov_rmp} on  the scaling of the defect density as a
function of rate of change of the Hamiltonian  following a slow quantum quench (see Refs.~[\onlinecite{aeppli10, dziarmaga_09, polkovnikov_rmp}] for a review), a series of works have been reported addressing the scaling behavior of the
 defect density for quench across a QCP, gapless phases, gapless  lines, etc \ct{mukherjee07}. Of a particular interest is generalization of these results to quenching
  across a multicritical point (MCP) \ct{divakaran09, mondal09, viola09,mukherjee10, viola10}. Such points (both for classical and quantum phase transitions) are characterized
 by two or more orthogonal directions with different correlation lengths~\cite{chaikin-lubensky}. They generically appear either as intersection of two critical lines
 or as a terminating point of one critical line on the other critical line. The second scenario is realized for the transverse field XY model, which is the primary
 example for this paper (see Fig.~\ref{fig:phase_diag}). In this model one terminating critical direction (horizontal line in Fig.~\ref{fig:phase_diag}) corresponds
 to a finite momentum anisotropic transition with the critical momentum vanishing at the MCP. In finite size systems the momentum is quantized in units of $1/L$.
 In turn this quantization results in emergence of quasi-critical points close to the MCP where the energy gap has a local minimum. As we will show
these quasi-critical points dominate the scaling of the fidelity susceptibility $\chi_{\rm F}$ along generic directions and in turn determine the scaling of
 defect density for slow as well as rapid quenches across a QCP as suggested
 in Refs.~\ct{gu09,gritsev09, grandi10}.

The scaling behavior of $\chi_{\rm F}$ at a QCP is well established \ct{venuti07,gurev08,shigu08}. For a Hamiltonian given by $H = H_0 + \lambda H_I = H_0 + \lambda V$ where $H_0$ is the Hamiltonian describing the QCP,  $V\equiv \partial_\lambda H\bigr|_{\lambda=0}=\sum_{r} v(r)$ is the perturbation not commuting with $H_0$  and $\lambda $ measures the deviation from the QCP. For the translationally invariant systems it can be shown that the fidelity susceptibility is related to the connected imaginary time correlation function of the perturbation~\cite{venuti07}:
\be
\chi_F(\lambda)=\int_0^\infty d\tau\int d{\bf r}\, \tau \langle v({\bf r},\tau) v(0,0)\rangle_c.
\ee
From this representation it is clear that the scaling dimension of $\chi_{\rm F}$ is ${\rm dim}[\chi_{\rm F}]=2\Delta_v-2 z -d$. If this scaling dimension is negative then the fidelity susceptibility diverges at the QCP as
$\chi_{\rm F} (\lambda=0) \sim L^{d + 2z - 2\Delta_v}$, where $L$ is the system size. If the scaling dimension is positive then the fidelity susceptibility is still singular but this singularity is a subleading correction on top of a nonuniversal constant (see e.g. Ref.~[\onlinecite{grandi10long}]). In case where $V$ is a marginal or relevant perturbation (so $\lambda v({\bf r})$ scales as the energy density) we have additional simplification coming from $\Delta_v=d+z-1/\nu$ so that at the critical point~\ct{schwandt09, gritsev09, capponi10, grandi10}
\be
\chi_{\rm F}\sim L^{2/\nu-d}.
\label{chif_1}
\ee
At small but finite $\lambda$ this scaling is substituted by
\be
\chi_{\rm F}\sim |\lambda| ^{\nu d -2}
\label{chif_2}
\ee
and the two asymptotes match when the correlation length $\xi\sim 1/|\lambda|^{\nu}$ becomes of the order of the system size. These asymptotics are dominant for $d\nu<2$ and subleading for $d\nu>2$ (at $d\nu=2$ there are additional logarithmic singularities~\cite{grandi10long}).

It is natural to expect that the scaling of the fidelity susceptibility can be further extended to understand universal properties of MCPs as well as quantum dynamics
 through the MCPs. For example, in Refs.~[\onlinecite{grandi10, grandi10long}] it was shown that the scaling of the generated quasiparticles and excess energy (heat)
 produced by slow quenches can be understood through the scaling behavior of generalized adiabatic fidelity susceptibilities. To the best of our knowledge such
 analysis was never performed before. As we mentioned earlier because MCPs are characterized by at least two orthogonal directions (principal axes) characterized
 by different scaling exponents, the natural object to analyze near the MCP is the so called geometric tensor~\cite{venuti07} defined as
\be
\chi_{\alpha\beta}=\sum_{n\neq 0} \langle 0|\partial_{\lambda_\alpha}|n\rangle\langle n|\partial_{\lambda_\beta}|0\rangle,
\ee
where the sum is taken over all excited states. One can also define the fidelity susceptibility, characterizing the drop of the overlap of the ground state wave functions for an infinitesimal change of the tuning parameter $\delta{\bm \lambda}$. It is clear that now $\chi_{F}$ is described by both the magnitude and the direction:
\be
\chi_{\rm F}=\sum_{\alpha,\beta} n_\alpha\, \chi_{\alpha\beta}\, n_{\beta},
\label{chi2}
\ee
where $n_\alpha$ is the projection of the vector $\delta\bm{\lambda}$ to the direction $\alpha$.  If we are interested in the fidelity susceptibility along a generic direction such that all components $n_\alpha$ are of the same order then the most divergent component of the geometric tensor defines the scaling of $\chi_{\rm F}$. It follows from the definition of the geometric tensor that its diagonal components describe the fidelity susceptibility along the principal axes. For the purposes of this work we will focus on the situation of two orthogonal directions $\alpha,\beta= 1,2$. It is straightforward to generalize our results to the  situation of multi-component MCPs.

It is natural to expect that both the geometric tensor and fidelity susceptibility will have singular behavior at the MCP. In this work we focus on how this singularity develops as we approach the MCP. We find that this approach to the MCP is highly nontrivial characterized by non-monotonic dependence of the components of the geometric tensor on the coupling ${\bm \lambda}$. In particular, in the thermodynamic limit
$L\gg \xi_1({\bm \lambda}), \xi_2({\bm \lambda})$, where $\xi_{1,2}({\bm \lambda})\sim |\lambda_{1,2}|^{-\nu_{1,2}}$, we find the that the geometric tensor can be expressed as:
\be
\chi_{\alpha,\beta}({\bm \lambda})\sim {|\lambda_1|^{d\nu_1}\over \lambda_\alpha\lambda_\beta} F_{\alpha,\beta} (|\lambda_2|^{\nu_2}/|\lambda_1|^{\nu_1}),
\label{chi_al_bet}
\ee
where $d$ is the dimensionality, $\nu_1$ and $\nu_2$ are the correlation length exponents corresponding to the couplings $\lambda_1$ and $\lambda_2$ respectively, and $F_{\alpha,\beta}$ is a scaling function.
 For approaching the MCP from a generic direction $\lambda_1=\lambda\sin\theta$ and $\lambda_2=\lambda\cos\theta$ with $\lambda\to 0$ and $\theta\neq 0,\pi/2,..$ fixed, the scaling of the geometric tensor is affected by the small argument asymptotic of the scaling function $F_{\alpha,\beta}(x)$ at $x\to 0$ (we assume that $\nu_1<\nu_2$). While for approach of the critical point along the line $|\lambda_1|^{\nu_1}\propto|\lambda_2|^{\nu_2}$ the argument of the scaling function is constant of the order of unity and the scaling of the geometric tensor is determined purely by the prefactor in Eq.~(\ref{chi_al_bet}). We point that some asymptotics of the scaling function in Eq.~(\ref{chi_al_bet}) can be constrained by critical exponents characterizing second order phase transitions away from the MCP. Thus if the $\lambda_2=0$ describes a line of critical points then the asymptotics of $\chi_{22}$ at $|\lambda_1|^{\nu_1}\gg |\lambda_2|^{\nu_2}$ should reduce to the scaling behavior of fidelity susceptibility of the corresponding phase transition (see Eq.~(\ref{chif_2})). Clearly this requirement determines the asymptotics of $F_{22}(x)$ at small $x$. One can find similar constraints on other components of $F_{\alpha,\beta}(x)$ at large and small $x$. In the next section we analyze the scaling functions and discuss their asymptotics for a particular one-dimensional transverse field XY-model.

In finite size systems the scaling (\ref{chi_al_bet}) is replaced by another expression:
\be
\chi_{\alpha,\beta}({\bm \lambda},L)\sim L^{1/\nu_\alpha+1/\nu_\beta-d}\tilde F_{\alpha,\beta}(L |\lambda_1|^{\nu_1}, L |\lambda_2|^{\nu_2}).
\label{chi_al_bet1}
\ee
If $L\gg 1/|\lambda_1|^{\nu_1}, 1/|\lambda_2|^{\nu_2}$ then the scaling above should reduce to Eq.~(\ref{chi_al_bet}). Exactly at the MCP ($\lambda=0$)
 the geometric tensor should be independent on $\lambda$ which leaves only two possibilities
 $\tilde F_{\alpha,\beta}(0,0)={\rm const}>0$ or $\tilde F_{\alpha,\beta}(0,0)=0$. Thus the nonzero components of the geometric tensor should have
 the scaling, which is a straightforward generalization of Eq.~(\ref{chif_1}). As it is clear from Eq.~(\ref{chi_al_bet1}) there is a third scaling
 regime where one argument of the scaling function $\tilde F$ is small and the other is large. For generic direction
 $\lambda_1\approx \lambda_2=\lambda$ this regime happens when the system size is bounded between the shorter and longer correlation lengths
 characterizing the system: $1/|\lambda|^{\nu_1}\lesssim L\lesssim 1/|\lambda^{\nu_2}|$ with a new scaling emerging when
 $L\approx \xi_1\sim 1/|\lambda|^{\nu_1}$. For the model we analyze we find that all components of the geometric tensor have additional singularity at
 this (quasi-critical) point leading to the local maximum of $\chi_{\alpha,\beta}$ with stronger scaling with $L$. We also find that in this intermediate
 regime $1/|\lambda|^{\nu_1}\lesssim L\lesssim 1/|\lambda|^{\nu_2}$ all components of the geometric tensor have strongly oscillatory dependence on the
 magnitude of $\lambda$ (or equivalently on the system size). These oscillations, however, maybe an artifact of our model where MCP corresponds to the
ending point of a nonzero momentum anisotropic transition. And finally at $L\sim \xi_2({\bm \lambda})$ the two scalings (\ref{chi_al_bet}) and
 (\ref{chi_al_bet1}) smoothly connect to each other.

The oscillations of the components of the geometric tensor and the fidelity susceptibility near the MCP in turn translate into a step-like behavior in the number of defects produced during a sudden quench of the magnitude $|{\bm \lambda}|$ starting from the critical point. The steps are more pronounced in relatively small systems, while in the limit of large system sizes we get a scaling behavior $n_{\rm ex} \sim  |\lambda|^{\nu_1 d}$. This result is similar to that observed in the rapid quench from a critical point~\ct{grandi10}.

\section{Geometric tensor for the transverse field XY-model}

\begin{figure}[htb]
\includegraphics[height=2.1in,width=2.8in, angle = 0]{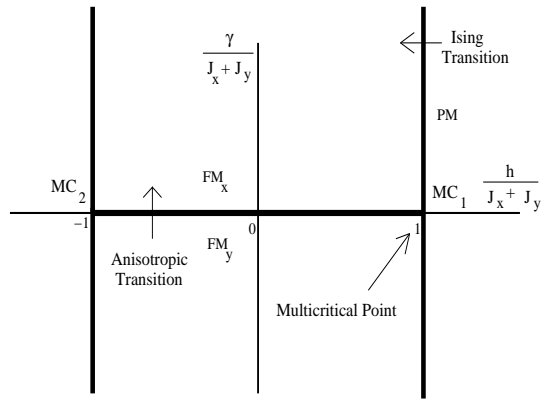}
\caption{The phase diagram of transverse $XY$ model. The vertical bold lines denote Ising transitions from the ferromagnetic phase to the paramagnetic phase , whereas  the horizontal bold line stands for anisotropic phase transition between two ferromagnetic phases. Critical exponents associated with both the Ising and the anisotropic transitions are $\nu=z=1$. The multicritical points (at $J_x=J_y$ and $h=\pm 1$) are denoted by $MC1$ and $MC2 $, respectively; the exponents are $\nu_1=1/2$, $\nu_2 = 1$ and $z_{mc}=2$. }
\label{fig:phase_diag}
\end{figure}

Let us begin with the spin-1/2 one-dimensional transverse XY Hamiltonian\ct{barouch70, bunder99, henkel, barouch71, aeppli10}
\ba
H = -\sum^{j = L}_{j = -L} \left({J_x \sigma^x_j \sigma^x_{j+1} + J_y \sigma^y_j \sigma^y_{j+1} + h\sigma^z_j}\right),
\label{ham_xy}
\ea
where $\sigma$'s are the Pauli spin matrices, $L$ is the system size, $J_x$, $J_y$ are the coupling constants along $x$
 and $y$ directions respectively, and $h$ denotes the transverse field. Here we consider even boundary condition ($\sigma^p_{L+1} = \sigma^p_L$,
 where $p = x,y,z$) in the odd fermion number sector, i.e., $\hat{P} = \frac{1}{2}\left(1 - \prod_{j=-L}^L {\sigma_j^z}\right) = 1$. As we will explain later
 choice of boundary condition does not affect the results in the thermodynamic limit. The boundary conditions will somewhat modify the finite size scalings (see below)
 but the exponents will remain unaffected. It is well known that the Hamiltonian (8) can be mapped to  a collection of non-interacting quasipartices, labeled by momentum modes $k$ ($k = 0, \pi/L, 2\pi/L, ... , 2\pi (L-1)/L$),  with an excitation energy of the  form \ct{lieb61, pfeuty70}
\ba
\Delta_k= \sqrt{(h + J\cos k)^2 + \gamma^2 \sin k^2},
\label{delta_k}
\ea
\label{delta_k}
where $J = (J_x + J_y)$ and $\gamma = (J_x - J_y)$. Clearly the spectrum is gapless along the line $\gamma=0$ provided that $-1\leq h/J\leq 1$ and along the two lines $h=\pm J$. At these boundaries the system undergoes continuous quantum phase transitions. The corresponding phase diagram is plotted in Fig.~\ref{fig:phase_diag}.
The points where the two critical lines meet are multicritical. We will focus on the MCP where $J=h$ and $\gamma=0$ (denoted by MC1 on Fig.~\ref{fig:phase_diag}).
 Then the vanishing gap corresponds to the momentum $k=\pi$ therefore it is convenient to shift momentum $k\to \pi-k$. Then the spectrum (9) linearized near the MCP simplifies to
\be
\Delta_k\approx \sqrt{(h-J+J k^2/2)^2+\gamma^2 k^2}
\label{delta_approx}
\ee
The ground state of the Hamiltonian (\ref{ham_xy}) is expressed as
\ba
|\psi_0 \ket = \prod_{k>0}{\left( \cos{\theta_k}|0_k,0_{-k}\ket + i\sin{\theta_k}|1_k,1_{-k}\ket \right)},
\ea
where the angles $\theta_k$ satisfy the relation\ct{gu09}
\ba
\tan {2\theta_k} &=& \frac{\gamma \sin k}{h + J\cos k} \non \\
&\approx& -\frac{\gamma k}{h + J(-1 + k^2/2)}.
\label{tan2theta1}
\ea
 Here the states $|0_k,0_{-k}\ket$, and $|1_k,1_{-k}\ket$ are respectively the empty and doubly occupied states of Jordan-Wigner fermions with momentum $k$ and $-k$.

\subsection{Scaling analysis of the spectrum near the MCP}

As we discussed earlier multicritical points are characterized by two (or more) tuning parameters~\ct{chaikin-lubensky}. This affects the scaling form of various observables. The natural tuning parameters in the present model are $\lambda_1= J_x + J_y - h$ and $\lambda_2=J_x-J_y = \gamma$. It is convenient to fix the overall energy scale $J\equiv J_x+J_y=1$. Then dispersion of the low energy  excitations (\ref{delta_approx}) can be rewritten in the scaling form
 \ba
 \Delta_k &\approx&  k^{z_{mc}} f\left(L|\lambda_1|^{\nu_1}, L|\lambda_2|^{\nu_2} \right),
 \label{delta_k1}
 \ea
where $z_{mc}=2$ is the dynamical exponent at the multicritical point, $\nu_1=1/2$ is the exponent assosciated with the divergence of the correlation  length when one approaches the MCP and $\nu_2$  is the correlation length exponent governing the cross-over from linear to quadratic dispersion at $\lambda_1=0$  (see also Ref.~[\onlinecite{damle96}]).
 Physically the exponent $\nu_2$ governs the scaling of the healing length determining the decay of the longitudinal correlations\ct{bunder99}.

The scaling function $f$ is explicitly written as follows:
\be
f(x_1,x_2)=\sqrt{\left(x_1^2-{1\over 2}\right)^2+x_2^2}
\label{sc_funct}
\ee
Because the MCP in our case is the intersection of two critical lines one of which terminates at the MCP $\lambda_2= 0$
 (the horizontal line on Fig.~\ref{fig:phase_diag}), the scaling function $f$ should satisfy some constraints. In particular, at $x_2\gg x_1,1$, the spectrum of
 excitation should reduce to that for the Ising critical point characterized by the exponent $z_{\rm it}=1$: $\Delta_k\sim k^{z_{\rm it}}$. This implies that
\be
f(x_1,x_2)\sim x_2^{\nu_2(z_{\rm mc}-z_{\rm it})}
\label{asympt1}
\ee
This is clearly the case since for $\nu_2=1$, $z_{\rm mc}=2$, and $z_{\rm it}=1$ we have $f(x_1,x_2)\sim x_2$ at large $x_2$. Likewise in the opposite limit of large $x_2\to 0$ the spectrum should be characteristic of the anisotropic transition occurring at finite value of $x_1\approx 1$ with a finite gap at $x_1=0$. This implies that for $x_2\ll x_1\ll 1$
\be
f(x_1,x_2)\sim {\rm const},
\ee
which is also the case. One can also infer the information about the scaling of the minimum gap near the anisotropic transition which occurs at $x_1=x_1^\star\approx 1$. Considering the limit $x_1=x_1^\star$ and $x_2\ll 1$ and noting the gap in the spectrum scales as $\lambda_2^{z_{\rm at}\nu_{\rm at}}$ from Eq.~(\ref{delta_k1}) we find that near the MCP the gap scales as
\be
\Delta_{\rm min}(\lambda_1,\lambda_2)\sim |\lambda_1|^{\nu_1 z_{mc}} \left|\lambda_2/\lambda_1^{\nu_1/\nu_2}\right|^{z_{\rm at}\nu_{\rm at}}.
\ee
Using the explicit exponents for our model we find that $\Delta_{\rm min}(\lambda_1,\lambda_2)\sim \sqrt{|\lambda_1|}|\lambda_2|$. For the generic direction where $\lambda_1\approx \lambda_2=\lambda$ this equation reduces to $\Delta_{\rm min}\sim |\lambda|^{3/2}$.  Similar analysis can reveal relations between correlation length exponents and the asymptotics of the scaling function.

\subsection{Scaling of the geometric tensor}

Next we will analyze the scaling of the different components of the geometric tensor. We will first perform the analysis specific for the transverse field XY model and then discuss generalizations to generic MCPs. Following eq. (4) and the definitions of $\lambda_1$, $\lambda_2$, we can rewrite Eq.~(\ref{tan2theta1})
\be
\tan(2\theta_k)\approx {\lambda_2 k\over \lambda_1-k^2/2}.
\label{tan2thetak}
\ee
It is easy to see that the components of the geometric tensor in our model take the form
\be
\chi_{\alpha\beta}={1\over L}\sum_{k>0} \langle 1k|\partial_{\lambda_\alpha}|0k\rangle\langle 0k|\partial_{\lambda_\beta}|1k\rangle,
\ee
where $|0k\rangle$ is the ground state wavefunction in the $k$ state and $|1k\rangle$ is the excited state of the pair of fermions with momenta $k,-k$:
\be
|1k\rangle=\sin\theta_k |0_k,0_{-k}\rangle-i \cos\theta_k |1_k,1_{-k}\rangle.
\ee
Using explicit properties of the wavefunction, this geometric tensor can be rewritten as
\ba
&&\chi_{\alpha\beta}={1\over L}\sum_{k>0} \left|\langle 1k|\partial_{\theta_k}|0k\rangle\right|^2 {\partial\theta_k\over\partial\lambda_\alpha} {\partial\theta_k\over\partial\lambda_\beta}={1\over L}\sum_{k>0} {\partial\theta_k\over\partial\lambda_\alpha} {\partial\theta_k\over\partial\lambda_\beta}\nonumber\\
&&={1\over 4L}\sum_{k>0} \cos^4 (2 \theta_k) {\partial \tan 2\theta_k\over\partial\lambda_\alpha} {\partial\tan 2\theta_k\over\partial\lambda_\beta}
\ea
With help of Eq.~(\ref{tan2thetak}) we can compute all three independent components of this tensor and find their scaling:
\ba
\chi_{11}&=&{1\over 4L}\sum_{k>0} {\lambda_2^2 k^2\over \Delta_k^4}; \quad \chi_{22}={1\over 4L}\sum_{k>0} {(\lambda_1-k^2/2)^2k^2\over \Delta_k^4};\nonumber\\
\quad \chi_{12}&=&-{1\over 4L}\sum_{k>0} {\lambda_2 k^2 (\lambda_1-k^2/2)\over\Delta_k^4}
\label{chi_munu}
\ea
Now, let us study each term of the tensor separately. First we focus on the MCP where $\lambda_1=\lambda_2=0$. In this case clearly the only nonzero component of the geometric tensor is $\chi_{22}$:
\be
\chi_{22}={1\over L}\sum_{k>0} {k^6\over k^8}={L\over 24}.
\label{chi_22}
\ee
Note that this scaling is exactly the same as across the Ising quantum critical points (vertical lines in Fig.~\ref{fig:phase_diag}). In this respect the geometric tensor evaluated right at the MCP does not contain any additional information about the additional singularity in the system compared to the standard critical point.

The situation changes dramatically if we consider the scaling of $\chi_{\alpha,\beta}$ close to the MCP. In particular, if we approach the MCP from a generic direction  $\lambda_1\approx \lambda_2$ with $\lambda_1>0$ (to be specific we assume $\lambda_1=\lambda_2=\lambda$)
 we see that the other diagonal component of the geometric  tensor $\chi_{11}$ becomes singular at the quasi-critical points occurring at $2\lambda = {k^*}^2 = (2\pi m)^2 L^{-2}$, with $m = 1,2,..$, where only one momentum mode $k^{*} = \sqrt{2\lambda}$ dominates the sum in Eq.~(\ref{chi_munu}). There we find
\be
\chi_{11}=\frac{1}{4L} \sum_{k>0}{\frac{\lambda_2^2 k^2}{\Delta_k^4}}\approx {1\over L} {(k^\star)^6\over(k^\star)^{12}}={1\over L (k^\star)^6}
\label{chi11a}
\ee
The lowest possible value of the dominant momentum for the quasi-critical point is $k^\star=2\pi/L$ (corresponding to $\lambda=2\pi^2/L^2$). At this coupling we find the maximum value for the $11$ component of the geometric tensor
\be
\chi_{11}^{\rm max}\approx {L^5\over (2\pi)^6}.
\label{chi11_max}
\ee
The next local maximum for $\chi_{11}$ will occur at the second quasi-critical point corresponding to $k^\star=4\pi/L$ and it is easy to see that it will be down by a
 factor of $2^6=64$ compared to the first maximum.  Once we go sufficiently far from the MCP all momenta will start contributing to the value of
 $\chi_{11}$ in Eq.~(\ref{chi_munu}) so that
\be
\chi_{11}(\lambda,\lambda)\approx \int\limits_{0}^{\infty} {dk\over 2\pi} {\lambda^2 k^2\over [(\lambda-k^2/2)^2+\lambda^2k^2]^2}\approx {\pi\over 4 \lambda^2}.
\label{chi11_av}
\ee
It is interesting that if we naively connect the scalings (\ref{chi11_max}) with (\ref{chi11_av}) we will get a mismatch. Indeed one can expect that when $\lambda$
 becomes comparable to $1/L^2$ the system size dependence in $\chi_{11}$ should disappear and the geometric tensor should become a function of $\lambda$ only with
 no $L$ dependence. This is indeed the case for usual critical points. However, here the situation is very different. At this value of $\lambda$ (corresponding to
 the shorter correlation length $\xi_1\sim {1/\sqrt{\lambda}}$ becoming comparable to the system size) the $11$ component of the geometric tensor starts to oscillate
 as a function of $\lambda$ and these oscillations persist until the moment when the contribution of the single  dominant momentum $k^\star$
  corresponding to the quasi-critical point becomes comparable with that of all other momentum modes. Comparing Eqs.~(\ref{chi11_max}) with (\ref{chi11_av}) we see that this happens when the second, longer,
 correlation length $\xi_2\sim 1/\lambda$ reaches the system size. So we conclude that highly oscillatory behavior of $\chi_{11}$ occurs in the intermediate values
 of the coupling $1/L^{1/\nu_1} < \lambda < 1/L^{1/\nu_2}$.

It is worthwhile to mention that in case of the Hamiltonian (8) with even boundary condition, the quantization of $k$ changes to  $k = (2m + 1)\pi/L$ in the $\hat{P} = 0$ sector \ct{henkel}. This shifts the position of the quasi-critical points by a factor of $1/L^2$, but does not change the $L^5$ scaling of $\chi_{11}^{\rm{max}}$. In the thermodynamic limit when the components of the geometric tensor are determined by the contributions from many momentum modes the boundary conditions become unimportant.

Similar analysis can be performed on the other two components of the geometric tensor. In particular, one can check that the off-diagonal component also has a very sharp singularity near the quasi-critical point rapidly changing from large positive to large negative value both of which scale as $L^4$.
 And finally the $\chi_{22}$ component also has two (positive) maxima on the left and on the right of the quasi-critical point which scale as $L^3$. Weaker singularities are observed in the vicinity of other quasi-critical points. As in the case of $\chi_{11}$ they persist until both correlation lengths become smaller than the system size.

Next we will attempt to generalize our scaling results to the case of generic MCP which has two different correlation length exponents $\nu_1$ and $\nu_2$. We will always assume that $\nu_1<\nu_2$ (when $\nu_1=\nu_2$ the scaling results reduce to those of a conventional QCP). For this purpose we will make the assumptions similar to those used in Refs.~[\onlinecite{polkovnikov05, grandi101}]. Namely, we will still rely on the assumption that elementary excitations are created in pairs with opposite momenta. In Refs.~[\onlinecite{gritsev09, grandi10long}] it was shown that this assumption is not affecting the universal scaling results, which alternatively can be obtained through the analysis of the correlation functions, but significantly simplifies the derivation. Besides in gapped systems lowest energy excitations generically have quasi-particle nature~\cite{sachdev99}. Then
\be
\chi_{\alpha\beta}=\sum_k \langle 0 |\partial_{\lambda_\alpha}| k\rangle \langle k |\partial_{\lambda_\beta}|0\rangle.
\label{chi_sum}
\ee
Here $|k\rangle$ denotes the state with two excited quasi-particles having momentum $k$ and $-k$ respectively.
Note that
\be
\langle 0 |\partial_{\lambda_\alpha}| k\rangle={1\over \lambda_\alpha} {\langle 0| \lambda_\alpha \sum_r v_\alpha(r) |k\rangle \over 2\Delta_k}.
\label{matr_el}
\ee
For relevant or marginal operators the scaling dimension of the perturbation $\lambda_\alpha\sum_r v_\alpha(r)$ is the same as the scaling dimension of energy (alternatively this perturbation is irrelevant). Therefore the scaling dimension of the matrix element (\ref{matr_el}) is the same as the scaling dimension of $1/\lambda_\alpha$. This can be expressed introducing the dimensionless scaling function
\be
\langle 0 |\partial_{\lambda_\alpha}| k\rangle={1\over\lambda_\alpha} f_\alpha (L|\lambda_1|^{\nu_1},L|\lambda_2|^{\nu_2}).
\label{matr_el1}
\ee
From this we can immediately infer the scaling of the geometric tensor in the thermodynamic limit when the summation over momenta can be substituted by integration:
\be
\chi_{\alpha,\beta}(\lambda_1,\lambda_2)={|\lambda_1|^{d \nu_1}\over\lambda_\alpha\lambda_\beta} F_{\alpha,\beta} (|\lambda_2|^{\nu_2}/|\lambda_1|^{\nu_1}),
\label{scale_chi}
\ee
where
\be
F_{\alpha,\beta}(x)=\int d^dq\, f_\alpha(q,q/x) f_\beta(q, q/x).
\label{g_al_bet}
\ee
Note that we implicitly assumed that the integral over momenta converges in the infrared limit. Otherwise one can expect nonuniversal cutoff dependent corrections to the geometric tensor coming from regularization~\cite{grandi10long}. For our particular XY model the diagonal components of the scaling function $F_{\alpha,\beta}$ are given by the following integrals:
\ba
F_{11}(x)&=&{1\over 4}\int_{0}^\infty {dq\over 2\pi} {q^2 x^2 \over \left[q^2x^2+(1-q^2/2)^2\right]^2},\\
F_{22}(x)&=&{1\over 4}\int_{0}^\infty {dq\over 2\pi} {q^2 x^2(1-q^2/2)^2 \over \left[q^2x^2+(1-q^2/2)^2\right]^2}.
\ea

If we would consider the curve $|\lambda_1|^{\nu_1}\sim |\lambda_2|^{\nu_2}$ then the argument of the scaling function $F_{\alpha,\beta}$ would be of the order of one and the scaling of the components of the geometric tensor will be given by the direct generalization of that for a QCP:
\be
\chi_{\alpha,\beta}\sim |\lambda_\alpha|^{d\nu_\alpha/2+d\nu_\beta/2-2}.
\label{scale_chi_1}
 \ee
However, this setup does not describe a generic situation. Because $\lambda_1$ and $\lambda_2$ are natural tuning parameters of the system it is more
 appropriate to analyze the scaling of the geometric tensor along the lines $\lambda_1=\lambda\cos\theta$ and $\lambda_2=\lambda\sin\theta$, where
 $\theta$ is some fixed slope. In this case it is easy to see that the argument of the scaling function $F_{\alpha,\beta}$ goes to zero as
 $\lambda\to 0$ because $\nu_1<\nu_2$. In turn this can affect the scaling of $\chi_{\alpha,\beta}$. In our case this is indeed the case and at
 small $x$ and $\cos\theta>0$ the integral in Eq.~(\ref{g_al_bet}) is determined by the interval $q\sim \sqrt{2}\pm x$, i.e. by the vicinity of the
 finite momentum critical points at $\lambda_2=0$ and $k=\sqrt{2\lambda_1}$.  The asymptotics of the functions $F_{\alpha,\beta}$ are then set by the
 properties of these finite momentum critical points for the anisotropic transition (see Fig.~\ref{fig:phase_diag}). E.g. the asymptotics of $F_{22}$
 is set from requiring that $\chi_{22}\sim |\lambda_2|^{d\nu_{\rm at}-2}$. Combining this with Eq.~(\ref{scale_chi}) we find that at small
 $x$ $F_{22}(x)\sim x$, which is indeed the case. Similarly from the fact that the fidelity susceptibility along $\lambda_1$ direction diverges
 as $1/|\lambda_2|$ at the anisotropic critical point we find that $F_{11}(x)\sim 1/x$ at small $x$. While these results are quite specific for the
 XY model the fact that they can be obtained from combining Eq.~(\ref{scale_chi}) with the critical properties of the transition occurring at
 $\lambda_2\to 0$ and finite $\lambda_1$ is generic.

For the path parametrized by $\lambda_1=\lambda\cos\theta$ and $\lambda_2=\lambda\sin\theta$ with $\cos\theta<0$ there are no singularities occurring at $\lambda_2\to 0$ because the critical line at $\lambda_2=0$ terminates at the MCP. In this case the scaling functions $F_{\alpha,\beta}(x)$ do not have any singularities at $x\to 0$ and we recover the regular scaling (\ref{scale_chi_1}).

And finally at very small values of $\lambda$ such that the sum in Eq.~(\ref{chi_sum}) is dominated by the single momentum $k^\star$ corresponding to the quasi-critical point we find that the scaling of the geometric tensor is determined by the scaling of the minimum gap at $k^\star$. Combining the scaling analysis with this observation we find that the maxima of $\chi_{\alpha,\beta}$ are determined by the scaling properties of the functions $f_\alpha$ evaluated at $k=k^\star$. In general using Eq.~(\ref{matr_el1}) and for simplicity assuming $d=1$ one can write the expression for the geometric tensor in the form (\ref{chi_al_bet1}) with
\ba
&&\tilde F_{\alpha,\beta}={L^{-1/\nu_\alpha-1/\nu_\beta} \over \lambda_\alpha\lambda_\beta}\times\nonumber\\
&&\sum_{n} f_\alpha \left({2\pi n\over L|\lambda_1|^{\nu_1}}, {2\pi n\over L|\lambda_2|^{\nu_2}}\right)f_\beta \left({2\pi n\over L|\lambda_1|^{\nu_1}}, {2\pi n\over L|\lambda_2|^{\nu_2}}\right).
\ea
For the system size smaller than the longer correlation length this sum can be dominated by a single most divergent term which determines the scaling of the geometric tensor. As we saw this is indeed the case for the transverse field XY model.

 Finally, we consider a path characterized by $\lambda_1 \approx \lambda_2^r$ for a general $r > 0$ (see appendix A). It can be shown \ct{mukherjee10} that
 quasi-critical points appear only for $r < r_c = 2$~ where $\chi_{11}$ diverges as  $\chi_{11} \sim L^{4/r + 1}$. In contrast, for $r \geq 2$, the scaling relation saturates to
 $\chi_{11} \sim L^{3}$ .

\subsection{Analysis of the geometric tensor through correlation functions.}

As we mentioned earlier the same scaling analysis can be repeated without making any assumptions about the nature of quasi-particles. Alternatively one can analyze the scaling of imaginary time correlation functions of the quench operator~\ct{venuti07, gurev08, li09}:
\be
\chi_{\alpha,\beta}(\lambda_1,\lambda_2)={1\over L^d}\int_0^\infty \tau G_{\alpha \beta}(\lambda_1,\lambda_2,\tau) d\tau,
\label{chi_munu1}
\ee
where
\ba
G_{\alpha \beta} (\tau) &=& \langle 0|\partial_{\lambda_\alpha} H(\tau)\partial_{\lambda_\beta}H(0)|0\rangle
\non \\ &-& \langle 0|\partial_{\lambda_\alpha} H(\tau)|0\rangle \langle 0| \partial_{\lambda_\beta}H(0)|0\rangle,
\ea
and
\be
\partial_{\lambda_\alpha} H(\tau) = e^{H \tau}\partial_{\lambda_\alpha} H e^{-H \tau}.
\ee
is the imaginary time Heisenberg representation of the operator $\partial_\lambda H$. Using our notations and translational invariance of the Hamiltonian we find
\be
\chi_{\alpha,\beta}(\lambda_1,\lambda_2)=\int_0^\infty d\tau \sum_r \langle 0|v_\alpha(r,\tau) v_\beta(0,0)\rangle_c,
\label{chi_munu1}
\ee
where the subscript ``c'' means that only the connected part of the correlation function should be taken into account. One can check that all the scaling results of the previous section can be reproduced through the language of the correlation functions. For the XY model we give the details of the derivation of the correlation functions in the Appendix (B). In particular, we find that at the quasi-critical point
\be
G_{11} (\tau) \sim \exp\left(L^{-3}\tau\right),
\ee
which again yields
\be
\chi_{11}^{\rm max} \sim L^5.
\ee

\subsection{Numerical analysis of the fidelity susceptibility along the fixed path.}

Having discussed the scaling of the geometric tensor, it is instructive to explicitly show the obtained dependencies. To be specific we will focus
on the fidelity susceptibility along a particular path given by $h = 2J_y$ fixed and $J_x$ being varied, which is equivalent to setting
$\lambda_1 = \lambda_2 = \lambda$. The system crosses the MCP ($J_x=J_y$ as shown in Fig.\ref{fig:phase_diag}).  As we discussed earlier the fidelity susceptibility is just the convolution of the geometric tensor with the derivatives $d\lambda_\alpha/d\lambda$ (see Eq.~(\ref{chi2})) so that in our case
\be
\chi_{\rm F}(\lambda)=\chi_{11}(\lambda,\lambda)+\chi_{22}(\lambda,\lambda)+2\Re\chi_{12}(\lambda,\lambda)
\ee
Clearly the fidelity susceptibility is dominated by the most divergent component of the geometric tensor.

\begin{figure}[htb]
\includegraphics[height=1.8in,width=2.4in, angle = 0]{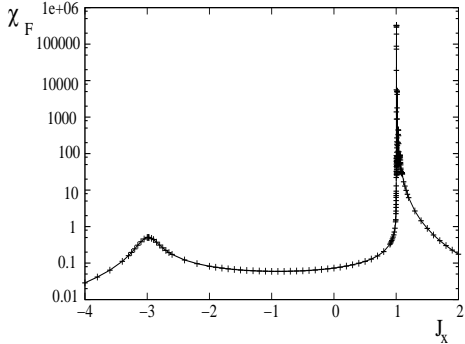}
\caption{The variation of $\chi_F$ with $J_x$, as obtained numerically using eq. (21), for a system size of $L = 100$. We have fixed $h = 2J_y = 2$.  The peak near the Ising critical point at $Jx = -3$ scales as $L$, whereas, the maxima near the MCP ($J_x = 1$) shows a $L^5$ divergence.}
\label{fig:chif_1}
\end{figure}

\begin{figure}[htb]
\includegraphics[height=1.8in,width=2.4in, angle = 0]{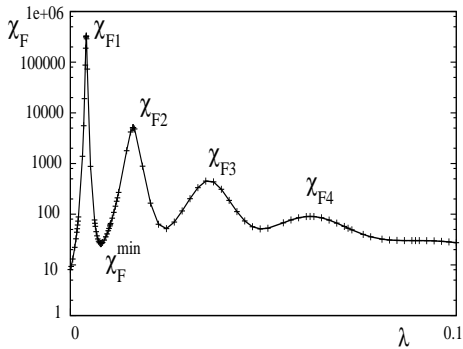}
\caption{Same as in Fig.~\ref{fig:chif_1} but near the multicritical point ($J_x = 1$).  The oscillatory behavior of the fidelity susceptibility is a signature of the presence of the quasi-critical points.  Each of the maxima, denoted by $\chi_{Fj},~j=1,2,..$, scales as $L^5$.}
 \label{fig:chif_2}
\end{figure}

\begin{figure}[htb]
\includegraphics[height=1.8in,width=2.4in, angle = 0]{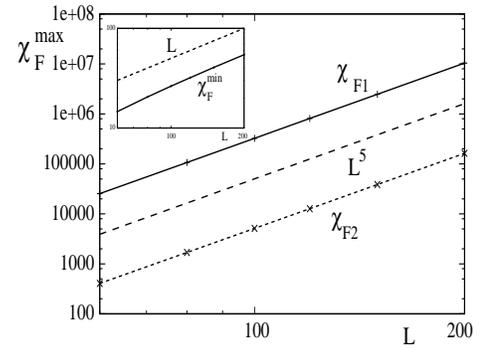}
\caption{The scaling behavior of $\chi_{\rm {F}}^{\rm max}$ with the system size $L$ at the first ($\chi_{\rm{F}1}$) and second ($\chi_{\rm {F}2}$) maxima near the
 multicritical point. The dashed line corresponds to the fixed slope $L^5$.
Inset: The first minima of $\chi_{\rm F}$, denoted by $\chi_{\rm F}^{\rm min}$ in Fig.~\ref{fig:chif_2}, shows a linear dependence on the system size.}.
\label{fig:chif_3}
\end{figure}

Combining Eqs.~(\ref{chi_munu}) we find
\ba
\chi_{\rm F} ={1\over L}\sum_{k > 0} \frac{(J_y \sin k  + 2J_y \sin k)^2}{\Delta_k^4}\approx {1\over 16 L}\sum_{k > 0}\frac{k^6}{\Delta_k^4}.
\ea
 Right at the MCP the scaling of $\chi_{\rm F}\sim L$ coincides with that of $\chi_{22}$ (see Eq.~(\ref{chi_22})). As $\lambda$ increases the fidelity susceptibility is becoming dominated by the $\chi_{11}$ component of the geometric tensor having very pronounced singularity $\chi_{\rm F}\sim L^5$ at the first quasi-critical point $\lambda=2\pi^2/L^2$.
In Figs.~\ref{fig:chif_1} and \ref{fig:chif_2} we show the numerically computed fidelity susceptibility for the XY-model. In Fig.~\ref{fig:chif_1} we focus on a 
larger range of $J_x$ which contains both the Ising critical point occurring at $J_x=-3$ (with the expected linear scaling of $\chi_{\rm F}$ with the system size) 
and the MCP occurring at $J_x=1$. While Fig.~\ref{fig:chif_2} focuses on the vicinity of the MCP. We see that indeed $\chi_{\rm F}$ has very pronounced non-monotonic 
behavior on the one side of the transition corresponding to $\lambda_1>0$. In Fig.~\ref{fig:chif_3} we show the scaling of the maxima with the system size and 
confirm $L^5$ asymptotics of the maximum of the fidelity susceptibility together with linear in $L$ scaling of the minima of $\chi_{\rm F}$. As we explained earlier
 the oscillatory behavior is due to proximity to the critical line corresponding to the anisotropic transition ($\lambda_2=0$) where critical points occur at finite 
momentum. Because $\nu_1=1/2<\nu_2=1$ the generic direction $\lambda_1=\lambda_2=\lambda$ is effectively pushed towards the anisotropic transition, 
indeed $|\lambda_1|^{\nu_1}\gg |\lambda_2|^{\nu_2}$. In turn this results in emergence of the quasi-critical points characterized by anomalously small gap scaling 
as $|\lambda|^{3/2}$ occurring when the allowed value of the momentum crosses the gapless point at the anisotropic transition 
line ($\lambda_1\approx (2\pi n/L)^2/2$) (see Fig.(5)).

\begin{figure}[htb]
\includegraphics[height=1.8in,width=2.4in, angle = 0]{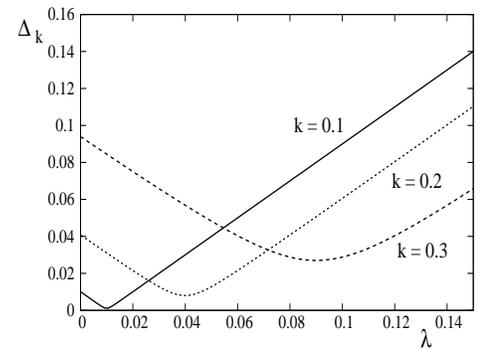}
\caption{{Gap $\Delta_k$  with $\lambda$ for the momentum modes $k = 0.1, 0.2$ and $0.3$, when $h = 2 J_y = 2$ is fixed, and
$\lambda_1 = \lambda_2 = \lambda = J_x - J_y$.
 The minimum of energy gap occurs at different values of $\lambda$ for different $k$, thus giving the quasi-critical points. }
}
\end{figure}

\subsection{Defect density following the instantaneous quantum quench from the MCP.}

\begin{figure}[htb]
\includegraphics[height=2.0in,width=2.6in, angle = 0]{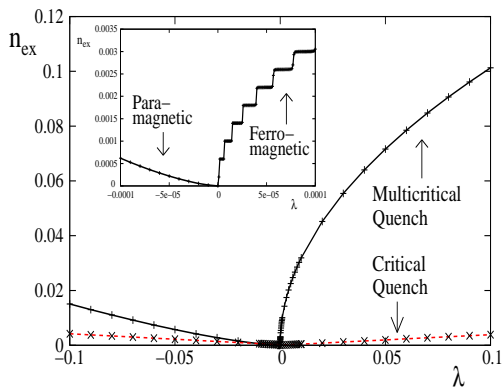}
\caption{(color online) Plot of kink density $n_{\rm ex}$ with the quench amplitude $\lambda$ for fast quench through critical (red line) and multicritical (black line) points, as obtained numerically, with a system size of $L = 5000$ spins, and $h = 2J_y = 2$.  Negative values of $\lambda$ correspond to quenching into the paramagnetic region, while positive values indicate the ferromagnetic phase. Inset: Sudden jumps in kink density for the same system size is shown to occur very close to the multicritical point, whenever $\lambda$ equals any  quasi-critical value. No such jumps are observed when system is suddenly quenched into the paramagnetic phase.}
\label{fig:n_ex}
\end{figure}

\begin{figure}[htb]
\includegraphics[height=1.9in,width=2.6in, angle = 0]{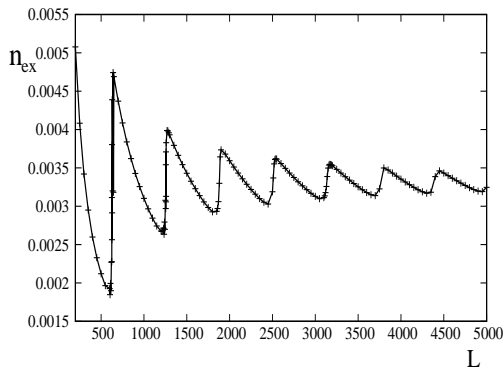}
\caption{Variation of kink density $n_{ex}$ with system size $L$ for fast quench through multicritical point, with $\lambda$ kept fixed at a very small value $0.0001$, and $h = 2J_y = 2$. }
\label{fig:n_ex1}
\end{figure}

\begin{figure}[htb]
\includegraphics[height=1.8in,width=2.5in, angle = 0]{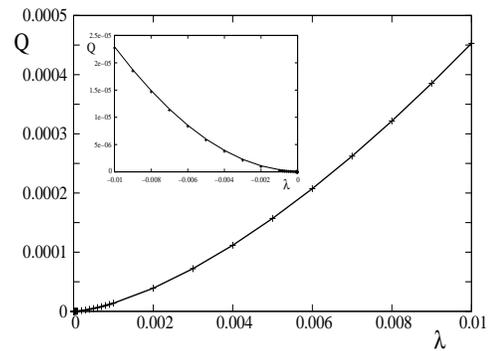}
\caption{ Plot of excitation energy $Q$ as a function of quench amplitude $\lambda$ for fast quench through the multicritical point into the ferromagnetic region,
 as obtained numerically, with a system size of $L = 5000$ spins, and $h = 2J_y = 2$ fixed. Inset: Plot of excitation energy $Q$ as a function of quench amplitude $\lambda$ for fast quench through the multicritical point
 into the paramagnetic region,}
\label{fig:Q}
\end{figure}

\begin{figure}[htb]
\includegraphics[height=1.8in,width=2.5in, angle = 0]{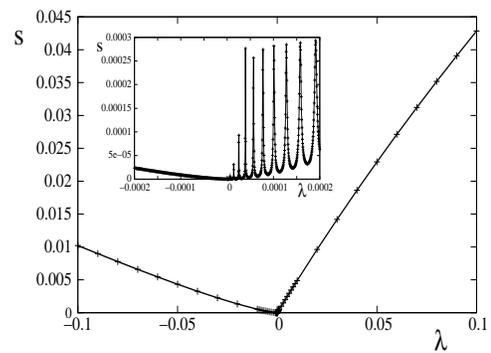}
\caption{ Plot of diagonal entropy density $s$ as a function of quench amplitude $\lambda$ for fast quench through the multicritical point,
 as obtained numerically, with a system size of $L = 5000$ spins, and $h = 2J_y = 2$ fixed. Negative values of $\lambda$ correspond to quenching into the paramagnetic region, while positive values indicate the ferromagnetic phase. Inset: Plot of diagonal entropy density for small quench amplitude $\lambda$.
Oscillations are observed near the quasicritical points inisde the ferromagnetic region.}
\label{fig:s}
\end{figure}

Recently it was understood that the density of defects (which is equivalent to the density of excited quasi-particles in the $XY$ model) produced during
 the quantum quench is closely related to the probability of exciting the system per unit volume, which in turn is related to fidelity~\ct{gritsev09, grandi10}.
 In the limit of large transverse field in the final Hamiltonian, when the ground state is all spins up (or down), the defect density becomes equivalent to the number of wrongly oriented spins.
 In the same spirit, we next look at the defect density ($n_{\rm ex}$) for a fast quench starting
from the MCP. For a usual critical point the defect density following a fast quench of magnitude $\lambda$ starting exactly at the QCP is given by~\ct{gritsev09, grandi10} (see also Fig.~\ref{fig:n_ex})
\ba
n_{\rm ex} \sim |\lambda|^{\nu d},
\label{nex1}
\ea
This scaling can be obtained from e.g. the adiabatic perturbation theory which states that (see Ref.~[\onlinecite{grandi10long}])
\be
n_{\rm ex}\approx \sum_k \left|\int_0^\lambda d\lambda' \langle 0|\partial_{\lambda'}|k\rangle\right|^2\leq
|\lambda|\int_0^\lambda d\lambda' \chi_{\rm F}(\lambda').
\ee
Then the scaling~(\ref{nex1}) immediately follows from that of $\chi_{\rm F}$. In our case the integral of the fidelity susceptibility diverges at small  $\lambda$ since $\chi_{\rm F}\sim 1/|\lambda|^2$ therefore we need to analyze the scaling of $n_{\rm ex}$ separately. Let us observe that as in the case of $\chi_{\rm F}$ the excitations are dominated by the matrix element $\langle 0|\partial_{\lambda_1}|k\rangle$ which scales as $1/|\lambda_1|$.  The second observation we make is that the sum over momenta $k$ is determined by the proximity to the anisotropic transition with shorter correlation length, i.e. $k\sim\sqrt{\lambda}$. Combining these considerations we find that
\be
n_{\rm ex}\sim |\lambda|^{d\nu_1}.
\ee
The other simpler way to get this scaling is to note again that for the generic quench where $\lambda_2\sim\lambda_1$ the spectrum is dominated by the
 direction with smaller value of $\nu$ (since then $|\lambda_1|^{\nu_1}\gg |\lambda_2|^{\nu_2}$. Thus the characteristic momentum above which
  transitions are suppressed is approximately equal to the inverse of the shorter correlation length $\tilde k\sim 1/|\lambda|^{\nu_1}$.
  This characteristic momentum (in power of dimensionality) determines the scaling of the defect density. The situation qualitatively changes for a
 quench with $\lambda_1<0$. In this case the spectrum is dominated by the critical line at $\lambda_2=0$ and we expect
 $n_{\rm ex}\sim |\lambda|^{d\nu_2}$.  Numerical analysis of fast quench into the paramagnetic
 region $\left(\lambda_1<0 \right)$ shows $n_{ex} \sim |\lambda|$ for large $|\lambda|$ ($|\lambda| \sim 0.1$). (see Fig. 6). The anisotropy of the MCP reveals itself, however, for the quenches with
 very small amplitude $\lambda\sim 1/L^2$, where only for the positive sign of $\lambda_1$ we observe step-like structure. These steps appear once a new
 quasi-critical point emerges when $\lambda\sim (2\pi m)^2/2L^2$, where $m = 1,2,3....L-1$. As a result, the  momentum mode $k_m = 2\pi m/L$ gets
 excited with high probability close to one leading to a sharp rise  of the order $1/L$ in the defect density. In contrast,
between any two consecutive quasi-critical  points there is no significant change in the total number of defects
(see the inset in Fig.~\ref{fig:n_ex}) and the  density does not appreciably change with $\lambda$. Same oscillations can be revealed by fixing the
 quench amplitude $\lambda$ at a very small value and varying the system size (see Fig.~\ref{fig:n_ex1}). There the non-monotonicity is related to the
 fact that the defect density decreases as $1/L$ with the system size between the jumps where the total number of excitations is approximately fixed.

We can extend the above analysis to the case of heat density \ct{gritsev09, grandi10} $Q$ (or the excess energy above the new ground state) which, unlike $n_{ex}$, can be defined even for non-integrable systems not describable in terms of independent quasi-particles. For a fast quench through MCP into the ferromagnetic region, $Q$ is expected to vary as
\ba
Q \sim \lambda^{\nu_1(d + z_{mc})}= \lambda^{1.5},
\ea
as is verified numerically (see Fig.~ \ref{fig:Q}). Further, in contrast to $n_{ex}$, $Q$ does not show abrupt rise at the quasi-critical points very close to
 the MCP, even though its slope with respect to $\lambda$ increases slightly at these points. This is a consequence of additional suppression of the
 energy by a factor of $\Delta(k^\star)\sim (k^\star)^3$.  A similar quench into the paramagnetic region gives
\ba
Q \sim \lambda^{2},
\ea
which is expected if we use the exponents $\nu=d=z=1$ relevant in this case.

We can also analyze the scaling of the diagonal entropy density \ct{grandi10,polkovnikov08}, 
\be
s =-\frac{1}{L}\left[\sum_{k>0} p_k\log(p_k)+ (1-p_k)\log(1-p_k)\right],
\ee
where $p_k$ is the probability to be in excited state of momentum $k$. Unlike $n_{\rm ex}$, the diagonal entropy density is defined even if we break integrability:
\be
s = -\frac{1}{L}\sum_n p_n\log(p_n)
\ee
where $p_n$ is the probability to occupy $n$-th eigenstate. The diagonal entropy is equivalent to von Neumann entropy of the time-averaged density matrix. For any $k$
mode, $s_k =-\left[p_k\log(p_k)+ (1-p_k)\log(1-p_k)\right]$ is close to zero for $p_k \approx 0$ (i.e., for $\lambda \lesssim k^2/2 $) or
 $p_k \approx 1$ (i.e., for $\lambda \gtrsim k^2/2 $), and assumes a finite value of order unity for $p_k \sim 1/2$. This results in oscillatory behavior of entropy
 density inside the ferromagnetic near the quasicritical points (see inset of Fig.~\ref{fig:s}). On the other hand, coarse grained diagonal entropy density inside the ferromagnetic 
region as well as the diagonal entropy density inside the paramagnetic region are dominated by $\nu = 1$, which gives
\ba
s \sim |\lambda|^{\nu d} \sim |\lambda|.
\ea 
In our present model, numerical results suggest $s \sim \lambda^{0.93}$ for quench into the ferromagnetic phase, whereas $s \sim \lambda^{1.25}$ for a fast quench into the paramagnetic region. 
(See Fig. \ref{fig:s}). 

On a related note, a recent study~\ct{mukherjee10} concerning slow non-linear quench across the MCP following a path parametrized by $\lambda_1 = -|\lambda_2|^r$ for a general $r > 0$ has shown defect density to
continuously vary with $r$ as $n_{ex} \sim \tau^{-r/6}$ ($1 \leq r \leq r_c = 2$) in the transverse field $XY$ model, where $\lambda_2$ varies with time $t$ as $\lambda_2 = t/\tau$, $-\infty < t < \infty$.
A similar behavior of fidelity susceptibility is also observed when we consider paths parametrized by $\lambda_1 \approx \lambda_2^r$, where $\chi_{11}^{\rm{max}}$ diverges as $\chi_{11}^{\rm{max}} \sim L^{4/r + 1}$ for $r \le r_c = 2$ (see appendix A).

\section{Conclusions.}
 We have studied the scaling of the geometric tensor and the fidelity  susceptibility  near a quantum MCP using the example of a spin-$\frac{1}{2}$ $XY$ chain in a
 transverse field. Our analysis shows that $\chi_{\rm F}$ oscillates near the MCP, with the peaks, given by $\chi^{\rm max}_F \sim L^5$, occurring at the
 quasi-critical points. In turn these points are associated with finite momentum anisotropic transition terminating at the MCP. The oscillations occur in the
 intermediate region of system sizes bounded between shorter and longer correlation lengths characterizing the vicinity of the MCP. Associated with these jumps we
 found a step-like behavior of generated defect density for sudden quenches of small amplitude starting from the MCP into the ferromagnetic region with a new jump occurring once a new
 quasi-critical point is crossed. In the thermodynamic limit these oscillations disappear and the defect density scales as $|\lambda|^{d\nu_1}$, where $\nu_1$ is a
 smaller correlation length exponent characterizing the MCP. A similar analysis of heat density gives a scaling of $\lambda^{\nu_1(d + z_{mc})}$, whereas
diagonal entropy density shows oscillations near the quasi-critical points, finally scaling as $s \sim \lambda^{\nu_2 d}$ in the limit of large $\lambda$. In contrast,
behavior inside the paramagnetic region is always dominated by the Ising critical line, with $\nu = z = 1$.

Using the adiabatic perturbation theory we generalized some of our findings to the situations of generic MCPs and showed how various scaling relations follow from
 the analysis of the scaling functions describing the transition matrix elements or equivalently the non-equal time correlation functions of the perturbation along
 the critical line characterized by a smaller correlation length exponent.

\acknowledgements
The authors acknowledge helpful discussions with V. Gritsev, M. Tomka, D. Sen and U. Divakaran. The work of A.P. was supported by NSF (DMR-0907039), AFOSR FA9550-10-1-0110, and Sloan Foundation.
A.D. acknowledges CSIR, New Delhi, for financial support. The authors also acknowledge Abdus Salam ICTP and SISSA, Trieste, for hospitality.

\appendix

\section{Appendix: scaling of the geometric tensor for a non-generic path}

In the main text we focused on the scaling of the geometric tensor along the generic direction $\lambda_1\approx \lambda_2$.
 While this setup is indeed most natural, it is instructive to analyze the scaling along other directions, in particular $\lambda_1= c \lambda_2^{r}$. For simplicity we choose
 $c=1$. This scheme lets us approach the multicritical point following a generalized path~\ct{mukherjee10}, parametrized by the exponent $r$.
 In Refs.~[\onlinecite{barankov08, sen08}] it was shown that for a nonlinear quench in time $t$ near a usual critical point where $\lambda\sim t^r$ the exponent $\nu_1$ associated with the amplitude $\lambda_1$ gets replaced by the product $r \nu_1$. A similar scenario occurs in our case. Indeed the scaling of gap can be written as
\ba
\Delta_k &\sim& k^{z_{mc}}f\left(\frac{|\lambda_1|}{k^{1/\nu_1}}, \frac{|\lambda_2|}{k^{1/\nu_2}} \right) \non \\
&\sim& k^{z_{mc}}f\left(\frac{|\lambda|}{k^{1/r\nu_1}}, \frac{|\lambda|}{k^{1/\nu_2}} \right),
\ea
where in the last step we have taken $\lambda_2 = \lambda_1^{1/r} = \lambda$. Similar transformations apply to the scaling of matrix elements.

For our setup of the $XY$ model in one dimension the spectrum gets modified to\ct{mukherjee10}
\be
\Delta_k \approx
k^{2} \sqrt{\left(\frac{\lambda^r}{k^{2}} - \frac{1}{2}\right)^2 + \left(\frac{\lambda}{k} \right)^2},
\ee
which gives\ct{mukherjee10}, for $r < r_c = \nu_2/\nu_1 = 2$, $\Delta_k \sim k^{2/r + 1}$ owing to the presence of quasi-critical points
 at $2\lambda = \left(2\pi m/L\right)^{2/r}$, $m = 1, 2, ...$. Whereas, for $r > r_c = 2$, quasi-critical points are absent, and the scaling saturates to $\Delta_k \sim k^2$. Repeating the same
 analysis as in the main text we find that the maxima of say $11$ component of the geometric tensor near the quasi-critical point now scale as
\ba
\chi_{11} = L^{\frac{4}{r} + 1}.
\ea
for $r < 2$. In the case of $r = 1$ we get back our previous result $\chi_{11} \sim L^5$. However, for $r \geq 2$
the exponent in the $L$-dependence of $\chi_{\alpha,\beta}$ saturates, oscillation disappear and we get
\ba
\chi_{11} \sim L^3,
\ea
which describes the scaling of the maximum value of the fidelity susceptibility for the Ising transition:
\be
\chi_{\rm F}^{\rm is}\sim {1\over L}\sum_k {\lambda^2\over k^2 (k^2+\lambda^2)^2}
\ee
occurring at $\lambda\approx k\approx 2\pi/L$. This critical power of $r_c=\nu_2/\nu_1$ separates the regimes of dominance of anisotropic and Ising transitions on the scaling of the geometric tensor.

\section{Evaluation of correlation functions}
The Hamiltonian $H$ at constant $J_1+J_2=1$ can be written as
\ba
H&=&-{1\over 2}\sum_j \left(\sigma^x_j \sigma^x_{j+1}+\sigma^y_j\sigma^y_{j+1}\right)\nonumber\\
&-&{\lambda_2\over 2}
\sum_j \left(\sigma^x_j \sigma^x_{j+1}-\sigma^y_j\sigma^y_{j+1}\right)-(1-\lambda_1)\sum_j\sigma^z_j ,
\ea
where $\lambda_1 \equiv 1 - h$ and $\lambda_2 \equiv \gamma$. Therefore,
\ba
&& \partial_{\lambda_1}H\equiv V_1 = \sum_j \sigma^z_j\nonumber\\
&&\partial_{\lambda_2}H\equiv V_2 =-{1\over 2}\sum_j \left(\sigma^x_j \sigma^x_{j+1}-\sigma^y_j\sigma^y_{j+1}\right).
\ea
Using translation invariance of the system we have \ct{venuti07,gurev08}
\be
\chi_{\alpha \beta}(\lambda_1,\lambda_2)= \int d\tau \sum_j \tau \langle V_\alpha(\tau,j) V_\beta(0,0)\rangle_c ,
\ee
where
\ba
\langle V_\alpha(\tau,j) V_\beta(0,0)\rangle_c &=& \langle V_\alpha(\tau,j) V_\beta(0,0)\rangle \non \\
&-& \langle V_\alpha(\tau,j)\rangle \langle V_\beta(0,0)\rangle.
\ea
In the above equation we have defined $A(\tau) = e^{H\tau} A e^{-H\tau}$ for any operator $A$.
As we have seen above, $\chi_{11}$ fixes the scaling of fidelity susceptibility at a quasi-critical point.
Hence let us concentrate on evaluating $\langle V_1(\tau,j) V_1(0,0)\rangle_c $. We have\ct{niemeijer67},
\ba
&& \langle V_1(j,\tau) V_1(0,0) \rangle_c = \langle\sigma_j^z(\tau) \sigma_0^z(0)\rangle - \langle\sigma_j^z(\tau)\rangle \langle \sigma_0^z(0)\rangle \non \\
&=& \langle\sigma_j^z(\tau) \sigma_0^z(0)\rangle - \langle m_z \rangle^2
= \left[\frac{1}{4\pi} \int^{\pi}_{-\pi} e^{ijk} e^{-\Delta_k \tau} dk\right]^2 \non \\
&-& \left[\frac{1}{4\pi} \int^{\pi}_{-\pi} e^{ijk}e^{-\Delta_k \tau}\cos{2 \Phi(k)} dk\right]^2 \non\\
&-& \left[\frac{1}{4\pi} \int^{\pi}_{-\pi} e^{ijk} e^{-\Delta_k \tau} \sin{2 \Phi(k)} dk\right]^2,
\ea
where $\Phi(k) = 1/2 \tan^{-1} \left[ \gamma \sin{k}/(h + \cos{k}) \right]$, and $\langle m_z \rangle$ is the magnetization along the $z$ direction. Taking into account that $\Delta_k$ and $\cos {2\Phi(k)}$ are even functions of $k$, and
$\sin{2\Phi(k)}$ is an odd function of $k$, we can rewrite eq. (B5) as
\ba
&& \langle V_1(j,\tau) V_1(0,0) \rangle_c = \left[\frac{1}{4\pi} \int^{\pi}_{-\pi}  e^{-\Delta_k \tau}\cos{jk} dk\right]^2 \non \\
&-& \left[\frac{1}{4\pi} \int^{\pi}_{-\pi} e^{-\Delta_k \tau}\cos{jk}\cos{2 \Phi(k)} dk\right]^2 \non\\
&+& \left[\frac{1}{4\pi} \int^{\pi}_{-\pi}  e^{-\Delta_k \tau} \sin{jk}\sin{2 \Phi(k)} dk\right]^2,
\ea

Now, $\cos jk \sim \sin jk \sim 1$. Let us consider the first term
\ba
I_1 = \left[\frac{1}{4\pi} \int^{\pi}_{-\pi} e^{-\Delta_k \tau} \cos jkdk \right]^2.
\ea
At a quasi-critical point, defined by $h + \cos{k^{*}} = 0$ and $\gamma = 1 - h$, the minimum value of $\Delta_k$ is $\Delta_{k^{*}} \sim {k^*}^3$, whereas, $\Delta_k \sim k^2$ for all the other momentum modes. Therefore
\ba
I_1 \sim \frac{1}{L^2} e^{-{k^{*}}^3 \tau},
\ea
where we have taken $dk \sim 1/L$ and $\cos{jk} \sim 1$.

Let us now concentrate on the second term
\ba
I_2 = \left[\frac{1}{4\pi} \int^{\pi}_{-\pi} e^{-\Delta_k \tau}\cos jk\cos{2 \Phi(k)} dk \right]^2.
\ea
At a quasi-critical point, $\gamma \sin{k^{*}}/(h + \cos{k^{*}}) \to -\infty$, which makes $\cos\left[{\tan^{-1} \left(\gamma \sin{k^{*}}/(h + \cos{k^{*}})\right]}\right) = \cos(-\pi/2) = 0$.
However, for the remaining $k$ modes, $\cos\left(\tan^{-1} \left[ \gamma \sin{k}/(h + \cos{k})\right]\right) \approx \cos(-k)\approx -1 + k^2/2$, and we have
\ba
I_2 \approx \left[\frac{1}{2L} \sum_{k\ne k^{*}} {e^{-k^2 \tau}} \cos kj\right]^2 \ll I_1
\ea
since $k^2 \gg k^3$.

Now, let us consider the third term
\ba
I_3 = \left[\frac{1}{4\pi} \int^{\pi}_{-\pi} { e^{-\Delta_k \tau}} \sin jk \sin{2 \Phi(k)} dk \right]^2.
\ea
Again, at a quasi-critical point, $\sin{2 \Phi(k^{*})} \sim 1$, whereas $\sin{2 \Phi(k)} \sim k$ for all $k \ne k^{*}$.
Therefore we can write $I_3$ as
\ba
I_3 \sim I_1 \sim \frac{1}{L^2} e^{-{k^{*}}^3 \tau}.
\ea

Therefore combining the above terms we get,
\ba
\langle V_1(j,\tau) V_1(0,0) \rangle_c \sim \frac{1}{L^2} e^{-{k^{*}}^3 \tau},
\ea
and
\ba
\chi_{11}^{\rm{max}} \sim \frac{1}{L} \int^{\infty}_{0} {\tau e^{-{k^{*}}^3 \tau} d\tau} \sim L^5.
\ea

\end{document}